\documentclass[prb,twocolumn,showpacs,superscriptaddress]{revtex4}
\usepackage{amssymb}

\usepackage{epsfig}
\usepackage{amsmath}

\begin{document}

\title{Exact ground state and elementary excitations of the spin tetrahedron chain}
\author{Shu Chen}
\affiliation {Department of Physics and Institute of Theoretical
Physics, The Chinese University of Hongkong, Hongkong, P. R.
China} \affiliation{ Beijing National Laboratory for Condensed
Matter Physics, Institute of Physics, Chinese Academy of Sciences,
Beijing 100080, P. R. China}
\author{Yupeng Wang}
\affiliation{ Beijing National Laboratory for Condensed Matter
Physics, Institute of Physics, Chinese Academy of Sciences,
Beijing 100080, P. R. China}
\author{W. Q. Ning}
\affiliation {Department of Physics and Institute of Theoretical
Physics, The Chinese University of Hongkong, Hongkong, P. R.
China} \affiliation{Department of Physics, Fudan University,
Shanghai 200433, P. R. China}
\author{Congjun Wu}
\affiliation{Kavli Institute for Theoretical Physics, University
of California, Santa Barbara, CA 93106, USA}
\author{H. Q. Lin}
\affiliation {Department of Physics and Institute of Theoretical
Physics, The Chinese University of Hongkong, Hongkong, P. R.
China}
\date{\today}

\begin{abstract}
We study the antiferromagnetic spin exchange models with $S=1/2$
and $S=1$ on a one-dimensional tetrahedron chain by both
analytical and numerical approaches. The system is shown to be
effectively mapped to a decoupled spin chain in the regime of
strong rung coupling, and a spin sawtooth lattice in the regime of
weak rung coupling with spin $2S$ on the top row and spin $S$ on
the lower row. The ground state for the homogeneous tetrahedron
chain is found to fall into in the regime of strong rung coupling.
As a result, the elementary excitation for the spin-$ 1/2$ system
is gapless whereas the excitation for the spin-$1$ system has a
finite spin gap. With the aid of the exact diagonalization method,
we determine the phase diagram numerically and find the existence
of an additional phase in the intermediate regime. This phase is
doubly degenerate and is characterized by an alternating
distribution of rung singlet and rung spin $2S$. We also show that
the $SU(3)$ exchange model on the same lattice has completely
different kind of ground state from that of its $SU(2)$
correspondence and calculate its ground state and
elementary excitation analytically. 
\end{abstract}
\maketitle

\draft

\section{Introduction}
The study of quantum antiferromagnetic spin models with strong
frustration has attracted great attentions over the past decades.
Early investigations of the frustrated quantum magnets were partly
motivated by the work of Anderson to search the
resonating-valence-bond (RVB) ground state in such systems
\cite{Anderson}. In the frustrated quantum magnets, the magnetic
ordering is generally suppressed by the frustration. Some
well-studied frustrated magnetic systems include, for example, the
kagome lattice and pyrochlore lattice, in which the interplay
between frustration and quantum fluctuation leads to rich
varieties of phenomena. Recently, the frustrated magnets are
believed to be prominent candidates of realizing spin liquid
states with exotic ground state and deconfined fractional
excitations \cite {Moessner,Fisher,Senthil2003}. While the
mechanism of deconfinment in two-dimensional (2D) magnetic systems
is less clear, the deconfinement of spinons in
quasi-one-dimensional magnetic system, which is closely related to
the phenomenon of spin charge separation, is well investigated
\cite {Faddeev,Affleck,Tsvelik,Gogolin}. Since 1980's, a large
number of low-dimensional frustrated magnets have been synthesized
experimentally\cite{JPC}. Generally speaking, strongly geometrical
frustration in these systems allows the simple dimerized state to
be the ground state of the low-dimensional frustrated spin system
and opens a spin gap. So far, a variety of quasi-one-dimensional
frustrated models have been studied theoretically\cite
{Majumdar70,SS,Lecheminant}.
Additionally, important
progress has been made in trapping cold atoms under a highly
controllable way very recently, and thus it stimulates intensive
investigation on how to simulate the magnetic systems using cold
atoms. A number of schemes have been proposed to implement a
variety of quantum spin models in optical lattices \cite{Duan}.

In this article, we investigate both the ground- and excited-state
properties of the spin models on a one-dimensional (1D)
tetrahedron chain as shown in Fig. 1, with site spins $S_i$
residing in four corners of each tetrahedron. The basic unit, i.e,
a tetrahedron is composed of four spins with equal
antiferromagnetic exchanges between each pair of spins. The
tetrahedron chain can be also viewed as a 1D pyrochlore strip in a
2D pyrochlore lattice, in which only two of four corners of each
tetrahedron are shared by a neighboring tetrahedron. Generally, a
three-dimensional (3D) pyrochlore lattice is a network of
corner-sharing tetrahedra and a 2D pyrochlore model, named also as
a checkerboard-lattice model, is obtained by a projection of the
3D lattice on a plane.  Different from the 1D pyrochlore strip
considering in this article, for both the 3D and 2D pyrochlore
lattices, each corner of the tetrahedron is shared by a
neighboring tetrahedron. As one of the most frustrated
antiferromagnets, the model of spin pyrochlore lattice has been
investigated by a variety of techniques including the
semiclassical large-$S$ limit, large-$N$ expansion of the $SU(N)$
model, the contractor renormalization method based on the cluster
expansion, and the bosonization method on the anisotropic
limit\cite{Canals,Sachdev04,Auerbach03,Starykh05}. In spite of the
intensive research, even the ground state properties of the 3D
pyrochlore lattice are not well understood. For the 2D pyrochlore
lattice, the numerical results based on exact diagonizations have
shown that the ground state has plaquette order\cite{Fouet}.
However, for the 1D pyrochlore strip, we can determine its ground
state and elementary excitation in an exact manner. With the aid
of numerical diagonalization of the corresponding spin lattice
systems with small sizes, we also investigate the quantum phase
transitions of the ground state due to the change of exchange
strengths along perpendicular rungs.

The spin model on a 1D pyrochlore strip as shown in Fig.1 is
described by the Hamiltonian:
\begin{equation}
H=J\sum_{\langle ij\rangle }\widehat{S}_i\cdot \widehat{S}_j,
\label{pyrochlore}
\end{equation}
where $\langle ij\rangle $ denotes sum over all the nearest
neighbors along the tetrahedron chain and $\widehat{S}_i$
represents the spin operator residing in site $i$.  In this work,
we study both the spin-$ 1/2 $ and spin-$1$ models on the
pyrochlore strip. Our results show that the excitation spectrum
for the spin-$1/2$ system is gapless and the elementary excitation
of the spin-$1$ system has a finite spin gap. This model can be
extended to the cases where the strengths of bonds among the
tetrahedra are not homogeneous. Here we only consider the
inhomogeneous case as shown in Fig. 1b, where we use $J_{\perp }$
to represent the exchange strength along the vertical rung of each
tetrahedron. For convenience, we rewrite the Hamiltonian of the
spin tetrahedron chain corresponding to Fig. 1b as
\begin{eqnarray}
H_A &=&J\sum_i\left[
\widehat{S}_{i,1}(\widehat{S}_{i,2}+\widehat{S}_{i,3})+(
\widehat{S}_{i,1}+\widehat{S}_{i,2}+\widehat{S}_{i,3})\widehat{S}%
_{i+1,1}\right]  \nonumber \\
&&+J_{\perp }\sum_i\widehat{S}_{i,2}\cdot \widehat{S}_{i,3}.
\label{pyrochloreA}
\end{eqnarray}
It is obvious that the model (\ref{pyrochloreA}) reduces to the
model (1) when $J_{\perp }=$ $J$, i.e., the homogeneous
tetrahedron chain (1) is a special case of the model
(\ref{pyrochloreA}). Since the model can be represented as a sum
of local Hamiltonian on each tetrahedron, a classical ground state
is obtained whenever the total spin in the tetrahedron is zero for
a homogeneous model. It is straightforward that the classical
ground states have a continuous local degeneracy.
\begin{figure}[tbp]
\includegraphics[width=3.2in]{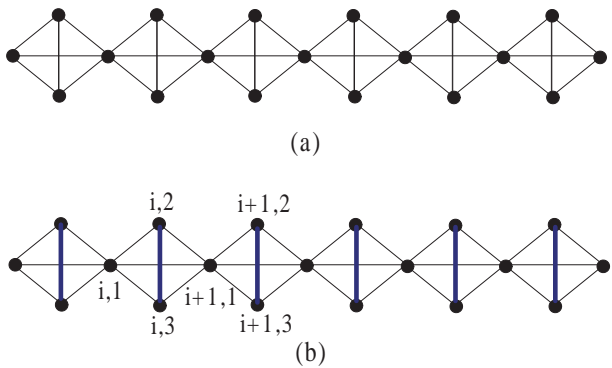}
\caption{(color online) Schematic pictures of (a) the homogeneous
spin tetrahedron chain, (b) the spin tetrahedron chain
corresponding to model (\ref{pyrochloreA}).} \label{fig1}
\end{figure}

\section{Spin-1/2 lattice}

Firstly, we consider the spin $1/2$ case. As we have mentioned in
the introduction, there is a fundamental different property
between the 1D pyrochlore strip and its high-dimensional
analogies: not all spins on a tetrahedron are equivalent. The
spins on the vertical rung of the tetrahedron are not coupled to
the neighboring tetrahedron. We define the total spin on each
vertical rung as
$\widehat{T}_i=\widehat{S}_{i,2}+\widehat{S}_{i,3}$. It is obvious
that $\widehat{T}_i^2$ is conserved. This property enables us to
simplify the 1D model greatly.

For the tetrahedron chain model (2), it is convenient to
reformulate Hamiltonian as
\begin{eqnarray}
H_A &=&J\sum_i\widehat{S}_{i,1}\widehat{S}_{i+1,1}+J\sum_i\left(
\widehat{S}
_{i,1}+\widehat{S}_{i+1,1}\right) \cdot \widehat{T}_i  \nonumber \\
&&+\frac{J_{\perp }}2\sum_i\widehat{T}_i^2-NJ_{\perp }S\left( S+1\right) ,
\label{PAA}
\end{eqnarray}
where $\widehat{T}_i^2=$ $T_i\left( T_i+1\right) $ with
$T_i=0,\cdots ,2S.$ In the strong coupling limit $J_{\perp
}\rightarrow \infty ,$ a pair of spins on each rung would form a
singlet (spin dimer) with $T_i=0$. This implies that the spins in
the horizontal direction along the chain is effectively decoupled
with the spins on the vertical rungs. Therefore, the ground state
of $H_A$ is a product of the ground state of spin chain and rung
singlets. Explicitly, it is represented as
\begin{equation}
\left| GS\right\rangle =\left| BA\right\rangle \bigotimes \prod_i\left[
S_{i,2},S_{i,3}\right]  \label{GSA}
\end{equation}
where $\left| BA\right\rangle $ denotes the Bethe-ansatz ground state
wavefunctions of the 1D Heisenberg chain $H_{chain}=J\sum_i\widehat{S}_{i,1}%
\widehat{S}_{i+1,1}$and $\left[ S_{i,2},S_{i,3}\right] =\left( \left[
\uparrow \right] _{i,2}\left[ \downarrow \right] _{i,3}-\left[ \downarrow
\right] _{i,2}\left[ \uparrow \right] _{i,3}\right) /\sqrt{2}$ is the dimer
singlet across the $ith$ vertical rung. The corresponding ground state
energy is
\begin{equation}
E_g^A=E_g^{BA}\left( N\right) -\frac 34NJ_{\perp },  \label{GEA}
\end{equation}
where $E_g^{BA}\left( N\right) $ is the ground state energy of the $N$-site
Heisenberg spin-$1/2$ chain. From the Bethe-ansatz solution of Heisenberg
model, we know the exact ground state energy $E_g^{BA}\left( N\right)
/N=-0.4431J$ at the infinite length limit.

In fact, utilizing the Raleigh-Ritz variational
principle\cite{SS,Sen96,Chen03,Han}, we can exactly prove the
state given by Eq. (\ref{GSA}) is the ground state of Hamiltonian
(\ref{pyrochloreA}) as long as $J_{\perp }\geq 2J.$ To see it
clearly, we can rewrite the Hamiltonian (2) with $J_{\perp }=2J$
as the sum of a Heisenberg chain and $2N$ projection operators,
which reads
\begin{eqnarray}
H_A
&=&J\sum_i\widehat{S}_{i,1}\widehat{S}_{i+1,1}+\sum_i\frac{3J}2\mathbf{P}
^{3/2}(\widehat{S}_{i,1},\widehat{S}_{i,2},\widehat{S}_{i,3})  \nonumber \\
&&+\sum_i\frac{3J}2\left[
\mathbf{P}^{3/2}(\widehat{S}_{i,2},\widehat{S}
_{i,3},\widehat{S}_{i+1,1})-1\right] ,
\end{eqnarray}
where
\begin{equation}
\mathbf{P}^{3/2}(\widehat{S}_{i,1},\widehat{S}_{i,2},\widehat{S}
_{i,3})=\frac 13\left[
(\widehat{S}_{i,1}+\widehat{S}_{i,2}+\widehat{S}_{i,3})^2-\frac
34\right]
\end{equation}
is a projection operator which projects a three-spin state
composed of $ S_{i,1},$ $S_{i,2}$ and $S_{i,3}$ into a subspace
with total spin $3/2.$ Now it is clear that the state given by Eq.
(\ref{GSA}) is the ground state of the global Hamiltonian because
it is simultaneously the ground state of each local
sub-Hamiltonian\cite{SS,Chen03,Han}. With the same reasoning, the
state (\ref{GSA}) is of course the ground state of Hamiltonian Eq.
(2) for $ J_{\perp }>2J.$ It is not hard to check that the state
(\ref{GSA}) is an eigenstate of the global Hamiltonian $H_A$ by
utilizing the identities $ \left(
\widehat{S}_{i,2}+\widehat{S}_{i,3}\right) \left[
S_{i,2},S_{i,3}\right] =0,$ however such an eigenstate is not
necessary the ground state for arbitrary $J_{\perp }.$ We note
that the condition $ J_{\perp }\geq 2J$ for the existence of the
ground state given by (\ref{GSA} ) is just a sufficient condition
which is a very strong restriction. In fact, it can be released to
a wider parameter regime. We expect that there is a critical value
$J_{\perp }^c$ and the system evolves into another quantum ground
state when $J_{\perp }$ is smaller than $J_{\perp }^c$.

For the homogeneous point $J_{\perp }=J$ which we are particularly
interested in, although the above proof is no longer applicable,
 we can still argue that Eq. (\ref{GSA}) remains the ground state, and
prove this result by using the numeric exact diagonalization
method. With the aid of numerical diagonalization, we may
determine the phase boundary of the model (\ref{pyrochloreA})
precisely. Since the total spin in every rung is conserved, the
vertical rungs are either in singlets or triplets for a spin-$1/2$
pyrochlore strip. Therefore, the relevant eigenstate of the
pyrochlore strip can be classified by the values of the total
spins on the vertical rungs\cite{Gelfand,Honecker00}. For
convenience, we use $E\left( N,M\right) $ to represent the
eigenenergy of the state with $N-M$ spin singlets and $M$ triplets
on the vertical rungs. It follows that the eigenenergy is given by
\begin{equation}
E\left( N,M\right) =-\frac 34NJ_{\perp }+MJ_{\perp
}+JE_{1/2,1}\left( N,M\right) , \label{ENM}
\end{equation}
where $E_{1/2,1}\left( N,M\right) $ represents the energy of the
lattice composed of $N$ spins with $S=1/2$ on the sites of the
lower row and $M$ spins with $T=1$ on the top row. For each class
of state with $N-M$ spin singlets and $M$ triplets on the vertical
rungs, there are altogether $C_N^M$ different configurations.
\begin{figure}[tbp]
\includegraphics[width=3.2in]{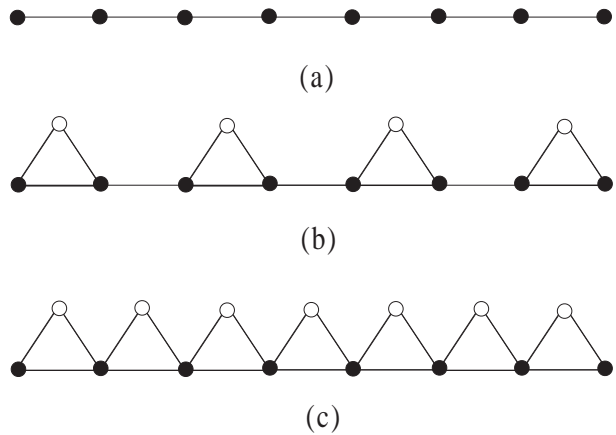}
\caption{Three types of ground state for the spin-$1/2$ and
spin-$1$ pyrochlore strips. The open and close circles denote the
spins with $S=1$ and $S=1/2$ respectively for the spin-$1/2$
pyrochlore strip. For spin-$1$ pyrochlore strip, the open circles
represent the spins with $S=2$ and the close circles represent the
spins with $S=1$.} \label{fig2}
\end{figure}

It is instructive to firstly consider the following two kinds of
configuration which correspond to two opposite limits of the
vertical exchange: (1) all the states on the vertical rungs are
spin singlets and (2) all the states on the vertical rungs are
spin triplets. In the first case, the original model can be mapped
to a spin chain model as displayed in Fig. 2a and the
corresponding eigenenergy is given by
\begin{equation}
E\left( N,0\right) =-\frac 34NJ_{\perp }+JE_{1/2}\left( N\right) ,
\end{equation}
where $E_{1/2}\left( N\right)=E_{1/2,1}\left( N,0\right) $ is the
energy of the $N$-site Heisenberg spin-$1/2$ chain and the
Bethe-ansatz ground state energy $E_{1/2}^g\left( N\right)
/N=-0.4431$ as $N\rightarrow \infty $ . In the second case, the
tetrahedron chain can be effectively described by a
$\bigtriangleup $-chain model consisted of $N$ spins with $S=1/2$
on the sites of the lower row and $N$ spins with $T=1$ on the
sites of top row as shown in Fig. 2c and the eigenenergy can be
represented as
\begin{equation}
E\left( N,N\right) =\frac 14NJ_{\perp }+JE_{1/2,1}\left( N,N\right) ,
\end{equation}
where $E_{1/2,1}\left( N,N\right) $ denotes the eigenenergy of the
corresponding $\bigtriangleup $-chain model. The mixed spin
$\bigtriangleup $ chain can be viewed as an alternating
spin-$1/2$-spin-$1$ chain with an additional next-nearest-neighbor
interaction between the spins with $S_i=1/2. $ It is well known
that there is a ferrimagnetic long-range order in the ground state
of quantum ferrimagnetic Heisenberg chain\cite
{Kolezhuk,Pati97,Wu99}. The additional interaction between the
spins with $ S=1/2$ is a kind of frustration which makes it harder
to compensate the spin with $S=1$, therefore the long-range order
still exists. Although no exact analytical results for the mixed
spin $\bigtriangleup $ chain are known, its ground state energy
may be determined by numerical exact diagonalization method. For a
mixed spin $\bigtriangleup $ chain with size of $8+8$, we get its
ground state energy given by $E_{1/2,1}^g\left( 8,8\right)
/16=-0.646773$. We note that the ground state of a spin
$\bigtriangleup $ chain or a spin sawtooth model with $S_i=1/2$ is
exactly known\cite{Sen96,Chen03,Blundell}.
\begin{figure}[tbp]
\includegraphics[width=3.2in]{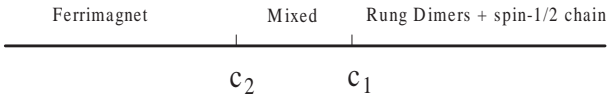}
\caption{The phase diagram for spin-$1/2$ tetrahedron chain with
variable vertical exchange $J_{\perp}$ in the parameter space of
$c=J_{\perp}/J$. For $c>c_1$, the system is in a decoupled phase
whose ground state is a product of rung singlets and the critical
spin liquid phase on the horizontal spin-$1/2$ chain; For
$c_2<c<c_1$, the ground state is a double degenerate mixed state
with the alternating spin singlet and spin triplet on the rungs;
For $c<c_2$, the ground state is a ferrimagnetic state with spin
triplet on the rungs.} \label{phasediagram1}
\end{figure}

It is clear that the state with all singlets or triplets on the
rungs are the ground state of the tetrahedron chain in the limit
of $J_{\perp } \rightarrow \infty $ and $J_{\perp }\rightarrow
-\infty$ respectively. Flipping a rung singlet into triplet costs
an energy of $J_{\perp }$, therefore the effect of the
antiferromagnetic coupling $J_{\perp }$ is to prevent the spins on
the rung forming triplet. On the other hand, a triplet in the rung
acts effectively as a spin with $S=1$ which lowers the total
energy by interacting with its neighboring spins with $S=1/2$ on
the lower rung. The competition between the two processes gives
rise to the complexity of the phase diagram for the tetrahedron
chain. An interesting question arising here is whether some
intermediate phases exist between the phases with fully paired
singlets and triplets on the rungs. To determine the phase
boundary of (\ref{pyrochloreA}) numerically, in principle, we need
consider all the different configurations of $T_i$ on the rungs.
Among a given class, we find that the configuration with the spins
on the top row repelling each other has lower energy. For example,
as shown in Fig. 2b, the configuration with alternating spin $1$
and spin $0$ on the top row has the lowest energy among the $
C_N^{N/2}$ configurations. After considering all the rung
configurations, we get a phase diagram as shown in the Fig. 3. As
expected, there is an intermediate phase which is effectively
described by the ground state of its equivalent model as shown in
Fig. 2b. Corresponding to Fig. 2b, there is another equivalent
configuration which is obtained by totally shifting the spins on
the top row a lattice space. In this phase, the triplet and
singlet on the rungs distribute in an alternating way and the
ground state is doubly degenerate.
\begin{figure}[tbp]
\includegraphics[width=3.2in]{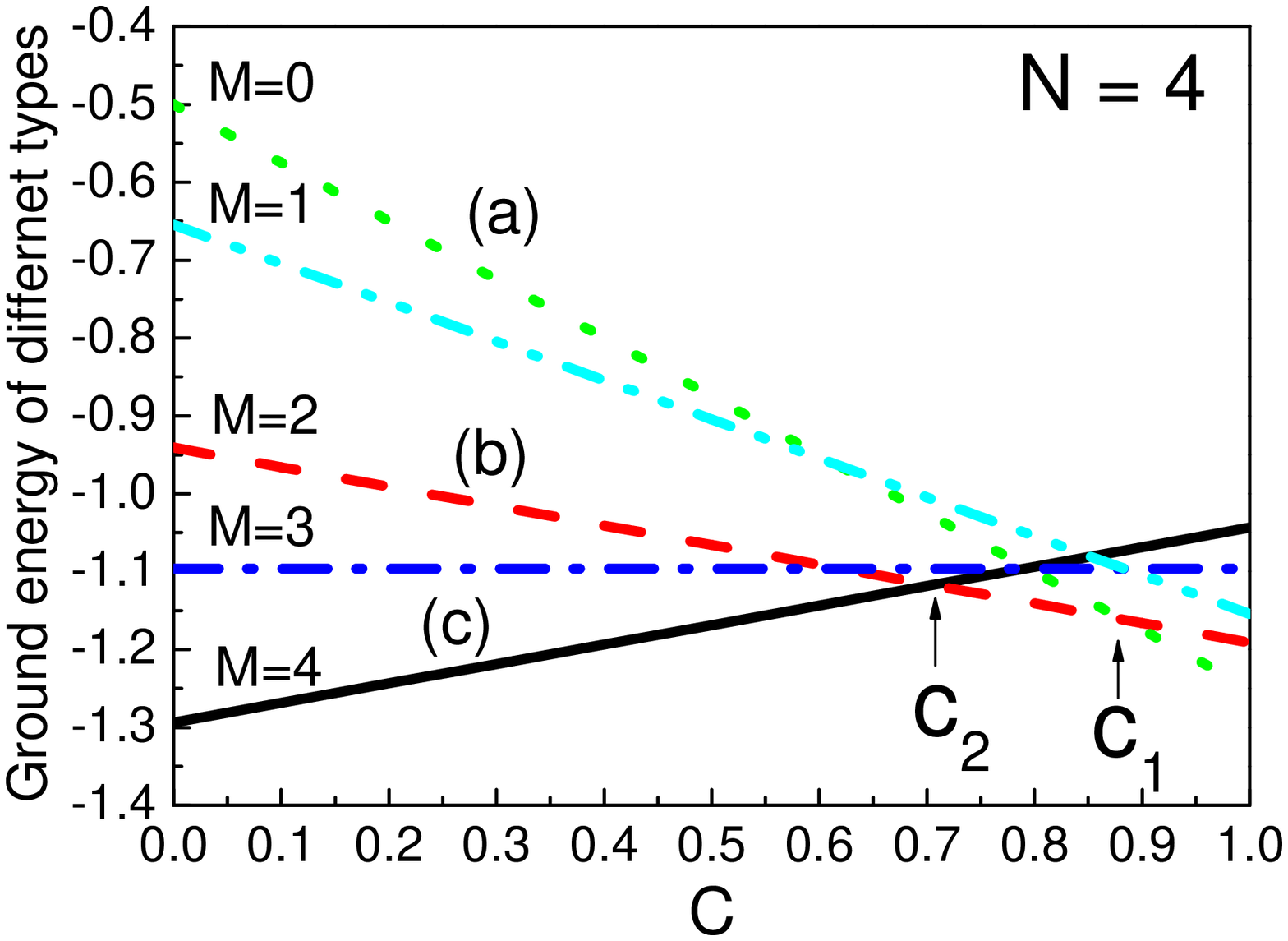}
\includegraphics[width=3.2in]{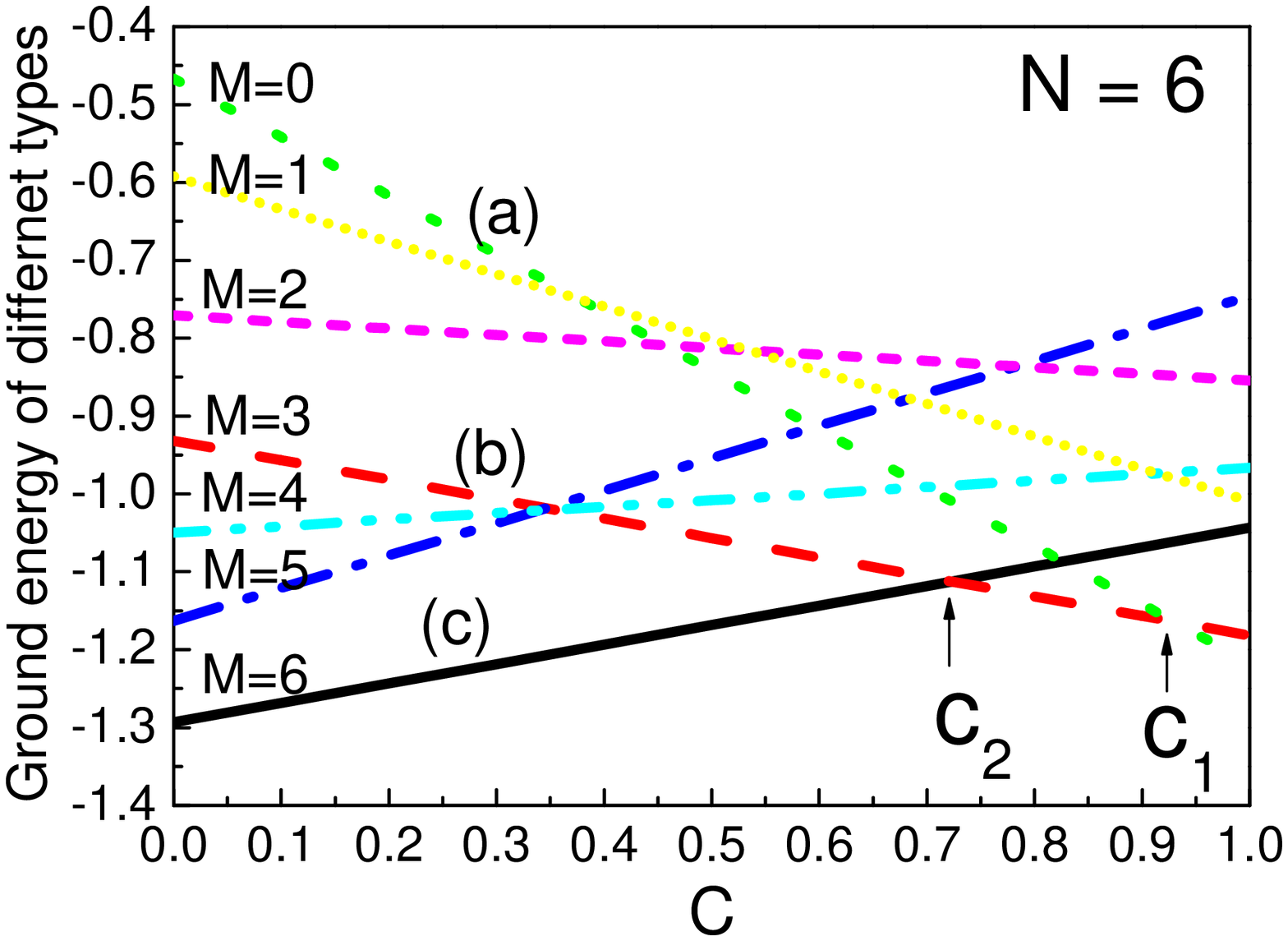}
\includegraphics[width=3.2in]{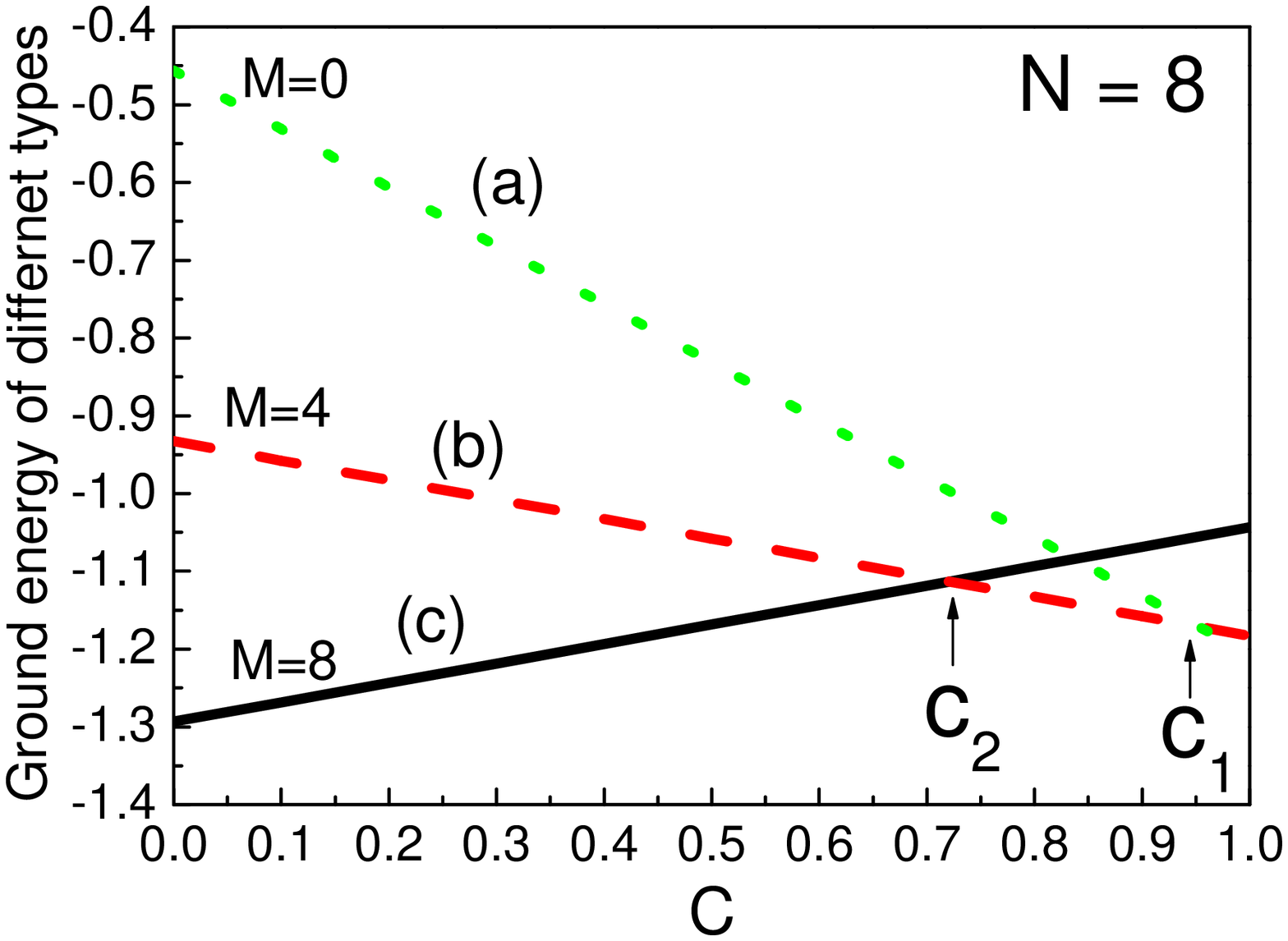}
\caption{(color online) The lowest energies of $E(N,M$)/J as a
function of $c=J_{\perp}/J$ for $N=4,6,8$. The solid, dashed and
dotted lines corresponding to the case with $M=N$, $M=N/2$ and
$M=0$ or (c), (b), (a) as displayed in Fig.2.}
\label{phasediagram1}
\end{figure}

Next we will give a description on how to determine the phase
diagram as shown in Fig.3.  Since the original pyrochlore strip
can be classified by the values of the total spins on the vertical
rungs or equivalently by $M$, labelling the numbers of the rung
triplets, the ground-state energy of our tetrahedral spin chain,
for a given value of J and $J_{\perp}$, will be given by the
lowest value of Eq. (\ref{ENM}) with $M=0,1,\cdots,N$. In Fig. 4,
we show as a function of $c=J_{\perp}/J$ all the lowest
eigenenergies for the lattice sizes with 4, 6, and 8 tetrahedra.
The energies corresponding to the point of $c=0$ (intercept points
with the vertical axis) are the lowest energies of
$E_{1/2,1}(N,M)$, from above to below, with $M=0,1,\cdots,N$. The
slope values of the straight lines depend upon the number $M$. Our
numerical analysis shows that the ground state in the whole
parameter space $c$ is determined by three kinds of configurations
with $M=0,N/2,N$, corresponding to the configurations of (a), (b)
and (c) displayed in Fig.2. and their crossing points determine
the phase transition points. For simplicity, we omit all the
excited energy levels for the case of $N=8$ in the Fig.4. From our
numerical results, we obtain the up critical value $ c_1=0.9529$
and the down critical value $c_2=0.7214$ for the original
pyrochlore strip with a size of $24$ sites (8 tetrahedra or
$N=8$). Similarly, we get $c_1=0.88186$ and $c_2=0.70546$ for
$N=4$,  $c_1=0.92952$ and $c_2=0.72351$ for $N=6$. Here $c_1$
corresponds to the crossing points of lines (a) and (b), whereas
$c_2$ corresponds to the crossing points of lines (b) and (c).
Since we need much smaller memory size to diagonalize systems of
(a) and (b) than system of (c), therefore we can calculate even
larger system to determine $c_1$. For example, we get
$c_1=0.96228$ for $N=10$ and $c_1=0.96752$ for $N=12$. In Fig. 5,
we analysis the finite size scaling of $c_1$ and $c_2$. The linear
fit of $c_1$ and $c_2$ gives $c_1=0.9767 \pm 0.0020$ and
$c_2=0.73021 \pm 0.00577$ for $N \rightarrow \infty$. We can also
determine the phase boundaries $c_1$ and $c_2$ in an alternative
way. By extrapolating the ground state energy of (a) and (b) to
infinite size, we then determine $c_1$ and $c_2$ by the crossing
of the energy levels of (a), (b) and (b), (c) respectively. In
Fig. 6, we display the finite size scaling of the ground state
energies of (a), (b) and (c). The ground state energies per site,
obtained by linear fit to the infinite limit, corresponding to the
configurations of (a), (b) and (c) are $0.44261\pm 0.00019$,
$0.62033\pm0.00091$ and $0.64675\pm0.00001$ respectively. By this
way and using Eq. (\ref{ENM}), we get $c_1=0.9757$ and
$c_2=0.72601$ in the limit of infinite size. If we use the ground
state energy obtained by Bethe ansatz for (a), we get
$c_1=0.9748$.
\begin{figure}[tbp]
\includegraphics[width=3.2in]{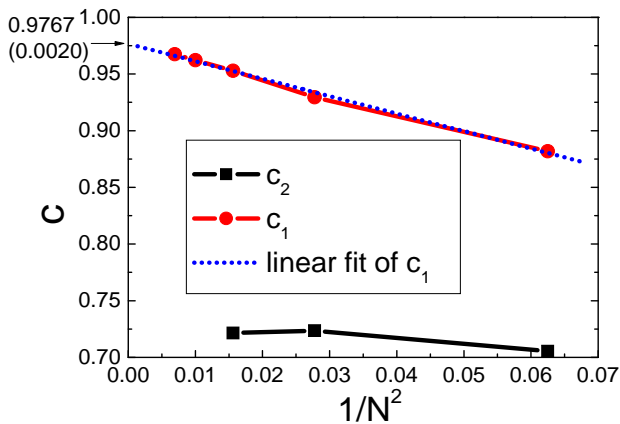}
\caption{(color online) The phase boundaries $c_1$ and $c_2$
versus the sizes of the system.}
\end{figure}

Before ending the discussion of the spin-$1/2$ pyrochlore strip,
we would like to give some remarks to a generalized spin
pyrochlore strip where the horizontal exchange is variable. With
the same reasoning as that of the model (\ref{pyrochloreA}), we
can easily get the sufficient condition for the existence of the
fully dimerized state on all the vertical rungs, which reads
$J_{\perp }\geq 2J$ and is irrelevant with the horizontal
exchange. When the horizontal exchanges are zero, the model has
totally different ground sates and it falls into the class of 1D
diamond mode\cite{Takano,Richter}, which has also fully dimerized
state on the vertical rungs as the ground state but the ground
state is highly degenerate with a degeneracy of $2^N$ because the
unpaired $N$ spins are completely free.
\begin{figure}[tbp]
\includegraphics[width=3.2in]{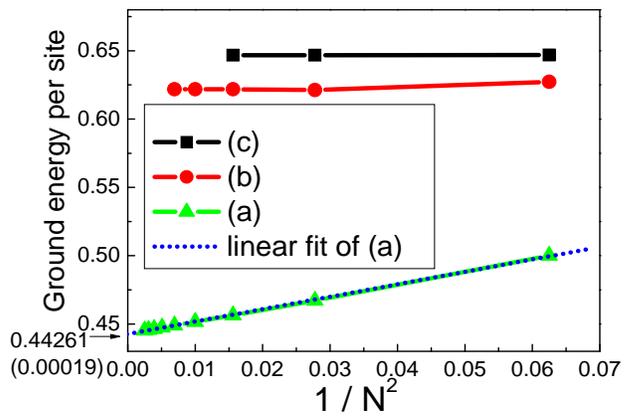}
\caption{(color online) The ground state energies of systems (a),
(b) and (c) versus the sizes of the system.}
\end{figure}

\section{Spin-1 lattice with SU(2) symmetry}

The above method can be directly extended to deal with the
spin-$1$ case. For the spin tetrahedron chain (\ref{pyrochloreA})
with $S=1,$ we can prove that the ground state of
(\ref{pyrochloreA}) is a product of fully dimerized singlets on
the rungs and the ground state of horizontal spin-$1$ chain as
long as $J_{\perp }\geq 4J$. Explicitly,the ground state can be
represented as
\begin{equation}
\left| GS\right\rangle =\left| Haldane\right\rangle \bigotimes \prod_i\left[
S_{i,2},S_{i,3}\right]  \label{GS_spin1}
\end{equation}
where $\left| Haldane\right\rangle $ denotes the ground state of the 1D spin-%
$1$ chain and
\[
\left[ S_{i,2},S_{i,3}\right] =\frac 1{\sqrt{3}}\sum_{m=-1}^1(-1)^{m+1}%
\left| m\right\rangle _{i,2}\left| -m\right\rangle _{i,3}
\]
is a spin singlet across the $ith$ vertical rung. The proof is rather
similar to its spin-$1/2$ correspondence and is straightforward when we
rewrite the Hamiltonian (\ref{pyrochloreA}) in the following form
\begin{eqnarray}
H_A &=&\sum_i\left[ \mathbf{h}_{\bigtriangleup }(\widehat{S}_{i,1},\widehat{S%
}_{i,2},\widehat{S}_{i,3})+\mathbf{h}_{\bigtriangleup }(\widehat{S}_{i,2},%
\widehat{S}_{i,3},\widehat{S}_{i+1,1})\right]  \nonumber \\
&&+J\sum_i\widehat{S}_{i,1}\widehat{S}_{i+1,1}
\end{eqnarray}
with $\mathbf{h}_{\bigtriangleup }(\widehat{S}_{i,1},\widehat{S}_{i,2},%
\widehat{S}_{i,3})=J\widehat{S}_{i,1}(\widehat{S}_{i,2}+\widehat{S}_{i,3})+%
\frac{J_{\perp }}2\widehat{S}_{i,2}\cdot \widehat{S}_{i,3}$ and $\mathbf{h}%
_{\bigtriangleup }(\widehat{S}_{i,2},\widehat{S}_{i,3},\widehat{S}%
_{i+1,1})=J(\widehat{S}_{i,2}+\widehat{S}_{i,3})\widehat{S}_{i+1,1}+\frac{%
J_{\perp }}2\widehat{S}_{i,2}\cdot \widehat{S}_{i,3}$. Now it is easy to
find that the ground state of $\mathbf{h}_{\bigtriangleup }$ is a product of
paired singlet on the rung and a free spin on the unpaired site as long as $%
J_{\perp }\geqslant 4J.$ By the variational principle, we conclude
that the state (\ref{GS_spin1}) is the ground state of spin-$1$
model given by Eq. (2) for $J_{\perp }\geqslant 4J.$ Certainly,
the sufficient condition for the existence of fully dimerized
ground state on the rung can be released to a lower bound. In
principle, we can determine it numerically following a similar
scheme of the spin-$1/2$ case.

\begin{figure}[tbp]
\includegraphics[width=3.2in]{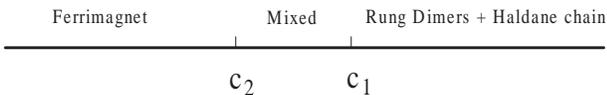}
\caption{The phase diagram for spin-$1$ tetrahedron chain. For
$c>c_1$, the system is in a decoupled phase whose ground state is
a product of rung singlets and Haldane phase on the horizontal
chain; For $c_2<c<c_1$, the ground state is a double degenerate
mixed state with the alternating spin singlet and spin quintet on
the rungs; For $c<c_2$, the ground state is a ferrimagnetic state
with spin quintet on the rungs.} \label{phasediagram2}
\end{figure}

Since the total spin $T_i$ on each vertical rung can be $0$, $1$
or $2$, the spin-1 tetrahedron chain can be mapped to a mixed
sawtooth-like model according to its configurations of the
vertical rungs. Several relevant configurations are: (1) Spin
singlets on all the vertical rungs with the eigenenergy given by
\begin{equation}
E_{T=0}=-2NJ_{\perp }+JE_1\left( N\right) ,
\end{equation}
where $E_1\left( N\right)$ denotes the eigenenergy of a spin-1
chain of N sites. From the known numerical results\cite{Lin90}, we
get ground state energy per site $E_{1}^g\left( N\right)
/N=-1.4051$ in the large $N$ limit. (2) Spin triplets on all the
vertical rungs with the eigenenergy given by
\begin{equation}
E_{T=1}=-NJ_{\perp }+JE_{1,1}\left( N,N\right) .
\end{equation}
Here $E_{1,1}\left( N,N\right) $ represents the eigenenergy of a
spin-1 $ \bigtriangleup $ chain of $2N$ sites\cite{Pati03}. (3)
Spin quintet $(T=2)$ on all the vertical rungs with the
eigenenergy given by
\begin{equation}
E_{T=2}=NJ_{\perp }+JE_{1,2}\left( N,N\right) .
\end{equation}
Here $E_{1,2}\left( N,N\right) $ represents the eigenenergy of a
mixed spin-1 and spin-2 $\bigtriangleup $ chain with spin $1$ on
the site of the lower row and spin $2$ on the site of top row.
Just similar to the spin-$1/2 $ pyrochlore strip, we need consider
all the different configurations of the rungs. Numerically, we
find the phase diagram is rather similar to that of the spin-$1/2$
case. As displayed in Fig. \ref{phasediagram2}, there is also an
intermediate phase between the fully paired singlet state with
$T_i=0$ and state with all $T_i=2.$ The intermediate state is
twofold degenerate with the configuration of alternating singlet
and quintet on rungs. Such a state can be schematized in terms of
Fig. 2b with the open circle denoting the state with $T_i=2$. From
our numerical results, we obtain $c_1=0.91596$ and  $c_2=0.88404 $
for the original tetrahedron chain with a size of $18$ sites (N=6)
as well as  $c_1=0.9134$ and $c_2=0.8452$ for $N=4$. With similar
reasoning as the spin-$1/2$ case, we can get $c_1=0.94189$ for
even larger system with $N=8$. Linear fitting of datum of $c_1$
versus $1/N^2$ and extrapolating it to the limit of the infinite
size, we get $c_1=0.95983 \pm 0.02748 $ for $N \rightarrow
\infty$.

As a natural generalization, it is straightforward to extend the
spin pyrochlore model with $SU(2)$ symmetry to the case with
arbitrary spin $S$. For the spin-$S$ model, a sufficient condition
for the existence of rung-dimerized ground state is $J_{\perp
}\geq 4-3/(s+1)$ for half-integer spin and $J_{\perp }\geq 4$ for
integer spin. In the dimerized phase, the horizontal chain is
decoupled with the spins on the rungs and therefore the elementary
excitation of the tetrahedron chain is gapless for
half-integer-spin model or opens a gap for the integer-spin model.

\section{Spin-1 lattice with SU(3) symmetry}

For the spin-1 system, the most general Hamiltonian has a biquadratic
exchange term besides the bilinear term and it exhibits much richer quantum
phase structures than the bilinear model\cite{Affleck89}.
When the
biquadratic exchange has the same strength of the bilinear exchange, the
Hamiltonian $H=J\sum_{\langle ij\rangle }h(i,j)$ with
\begin{equation}
h(\widehat{S}_i,\widehat{S}_j)=\widehat{S}_i\cdot
\widehat{S}_j+(\widehat{S}_i\cdot \widehat{S}_j)^2
\end{equation}
owns the $SU(3)$ symmetry. For spin-1 systems of transition metal
compounds where two electrons are coupled ferromagnetically by
Hund's rule, the biquadratic exchange term originates from a
fourth order perturbation process. Its magnitude is thus small
compared to the bilinear terms, and thus the $SU(3)$ symmetry is
not applicable. However, in the cold atomic physics, most atoms
have high hyperfine multiplets, thus it is possible to achieve
high symmetries. For example, the $^6$Li atom is with nuclear spin
1 and electron spin 1/2. In a weak magnetic field, electron spin
is polarized, while nuclear spin remains free. Recent studies
indicated that the three nuclear spin components can be described
by an approximate $SU(3)$ symmetry \cite{Honerkamp2004}.

It is well known that the $SU(3)$ exchange model on a chain has
quite different properties from that of the $SU(2)$ bilinear
model. Therefore we may expect that the $SU(3)$ tetrahedron chain
\begin{eqnarray}
H &=&J_{\perp
}\sum_ih(\widehat{S}_{i,2},\widehat{S}_{i,3})+J\sum_i\sum_{
\alpha =1}^3h(\widehat{S}_{i,\alpha },\widehat{S}_{i+1,1})+  \nonumber \\
&&J\sum_i\left[
h(\widehat{S}_{i,1},\widehat{S}_{i,2})+h(\widehat{S}_{i,1},
\widehat{S}_{i,3})\right]  \label{HSU3}
\end{eqnarray}
also displays different phase structure from that of its $SU(2)$
correspondence (\ref{pyrochloreA}) . The model with $J_{\perp }=J$
was initially proposed by three of us with S. C. Zhang in Ref.
\cite{Chen05} as an example of the $SU(N)$ generalization of the
Majumdar-Gosh model, however no analytical results have been given
there. Observing that the Hamiltonian can be written as a sum of
the Casimir of the total spin in each tetrahedron and the
representations with the smallest Casimir made out of four sites
in the fundamental representations is three dimensional, we
concluded that the state of trimer products is the GS of the
$SU(3)$ tetrahedron chain. The ground state is twofold degenerate.
Explicitly, the ground state of the $SU(3)$ spin tetrahedron chain
can be represented as
\begin{equation}
\left| GS\right\rangle _1=\prod_iT\left( S_{i,1},S_{i,2},S_{i,3}\right)
\end{equation}
or
\begin{equation}
\left| GS\right\rangle _2=\prod_iT\left( S_{i,2},S_{i,3},S_{i+1,1}\right) ,
\end{equation}
where
\[
T\left( S_i,S_j,S_k\right) =\frac 1{\sqrt{6}}\epsilon _{\alpha \beta \gamma
}\left| \alpha _i,\beta _j,\gamma _k\right\rangle
\]
represents a trimer state which is a singlet composed of three spins on site
$i,$ $j,$ and $k$ . Here $\alpha _i$ denotes the spin on site $i$ with the
value $\alpha $ taking $1,$ $0,$ or $-1$ and $\epsilon _{\alpha \beta \gamma
}$ is an antisymmetric tensor$.$

In the following, we shall calculate the ground state energy and elementary
excitation of Eq. (\ref{HSU3} ) analytically. For convenience, we make a
shift of constant to the Hamiltonian (\ref{HSU3} ) by replacing $h(\widehat{S%
}_i,\widehat{S}_j)$ with $\widehat{P}_{i,j}=h(\widehat{S}_i,\widehat{S}%
_j)-1. $ We note that modification of $J_{\perp }$ does not lift
the degeneracy of the left- and right-trimer states. For $J_{\perp
}>J,$ the state of trimer products is of course the ground state
of the SU(3) tetrahedron chain and the corresponding ground state
energy is
\[
E_g=-2NJ-NJ_{\perp }.
\]
Breaking a singlet of trimer will cost a finite energy, thus the
elementary excitation of the SU(3) tetrahedron chain has an energy
gap. For a three-site cluster, the trimer singlet is represented
by a Young tableaux $ [1^3]$ and the first excited state above the
singlet are represented by the Young tableaux $[2^11]$. When a
trimer singlet is broken, it decomposed into a monomer and a
paired dimer. For a system with degenerate ground state, the
monomer and dimer can propagate freely in the background of
trimerized ground state and lower the energy further. In
principle, two type of excitations are available in a pyrochlore
chain, either a magnon-like excitation produced by flipping a
trimer state into its excited state or a pair of deconfined
objects composed of a dimer plus a monomer. For our system with
doubly degenerate ground state, the spinon-like excitations have
lower energy.

The deconfined excitations behave like domain-wall solitons which connects
two spontaneously trimerized ground states. Explicitly, we represent an
excited state with a dimer at site 2m-1 and a monomer at site 2n is
represented as
\begin{eqnarray*}
&&\Psi \left( m,n\right)  \\
&=&\cdots T\left( S_{m-1,1},S_{m-1,2},S_{m-1,3}\right) m(S_{m,1}) \\
&&T\left( S_{m,2},S_{m,3},S_{m+1,1}\right) \cdots T\left(
S_{n-1,2},S_{n-1,3},S_{n,1}\right)  \\
&&d\left( S_{n,2},S_{n,3}\right) T\left(
S_{n+1,1},S_{n+1,2},S_{n+1,3}\right) \cdots ,
\end{eqnarray*}
where $d\left( S_i,S_j\right) =\frac 1{\sqrt{2}}\left( \left| \alpha _i\beta
_j\right\rangle -\left| \beta _i\alpha _j\right\rangle \right) $ with $%
\alpha \neq \beta $ represents a dimer. The corresponding momentum-space
wavefunction is
\[
\Psi \left( k_m,k_d\right) =\sum_{1\leq m\leq n\leq M}e^{imk_m+ink_d}\Psi
\left( m,n\right) .
\]
The excitation spectrum can be calculated directly by using the above
variational wavefunction. Because there exists no intrinsic mechanics
responsible for binding the dimer and monomer together to form a bound state
in a spontaneously trimerized system, it is reasonable to assume that the
dimer and monomer are well separated and they could be treated separately.
Similar schemes have been used to evaluate the excitation spectrum in the
spin-sawtooth system\cite{Sen96,Chen03}. Under such an approximation, the
excitation spectrum can be represented as a sum of monomer part and dimer
part, i.e., $\omega \left( k_m,k_d\right) \approx \omega \left( k_m\right)
+\omega \left( k_d\right) .$ The state $\Psi \left( n\right) $ is not
orthogonal with the inner product given by
\[
\left\langle \Psi \left( n^{\prime }\right) \right| \left| \Psi \left(
n\right) \right\rangle =\left( \frac 13\right) ^{\left| n^{\prime }-n\right|
},
\]
thus $\Psi \left( k_d\right) $ has a nontrivial norm $\left\langle \Psi
\left( k_d\right) \right| \left| \Psi \left( k_d\right) \right\rangle
=4/(5-3\cos k_d)$. With a similar scheme as solving the spectrum of
excitations of the spin-$1/2$ model \cite{SS,Sen96,Chen03}, we get the
spectrum for the dimer
\begin{equation}
\omega \left( k_d\right) =\frac 54\Delta _d-\frac{3\cos k_d}4\Delta _d
\end{equation}
with $\Delta _d$ $=2J$. In the large $J_{\perp }$ limit, a monomer can only
hop around in the left and right phases along the horizontal direction
without breaking additional trimer singlet, therefore its excitation
spectrum form a flat band, i.e., $\omega \left( k_m\right) =0$. When $%
J_{\perp }$ is close to $J,$ the monomer actually can move around
several corners, i.e., the monomer can be in the site of $\left(
m,2\right) $ and $ \left( m,3\right) $ either, therefore the
wavefunction for a monomer has resonating structure at the $m$th
triangle in the pyrochlore. The process of hopping from site
$\left( m,1\right) $ to $\left( m,2\right) $ or $\left( m,3\right)
$ accompanies with an energy cost of $J_{\perp }-J$, and thus the
excitation spectrum is still a flat band for the homogeneous
pyrochlore with $J_{\perp }=J$. For the inhomogeneous case, we take
the monomer excitation as a three-site cluster consisted of three
single monomers at sites $\left( m,1\right) $, $\left( m,2\right)
$ and $\left( m,3\right) $, i.e.,
\begin{eqnarray*}
\Psi \left( m\right)  &=&\frac 1{\sqrt{3}}\left[ \cdots m(S_{m,1})\cdots
+\cdots m(S_{m,2})\cdots \right.  \\
&&\left. +\cdots m(S_{m,3})\cdots \right] .
\end{eqnarray*}
Similarly,  after considerable algebra, we get the spectrum for
the monomer
\begin{equation}
\omega \left( k_m\right) =\frac{2}{3}\Delta _m-\frac{6\cos
k_m}{15}\Delta _m
\end{equation}
with $\Delta _m$ $=J_{\perp }-J$.

\section{Conclusions}

We have proposed and studied a class of frustrated lattice which
can be viewed as a 1D strip of the pyrochlore lattice or a
tetrahedron chain. For the general Heisenberg exchange model, we
give an exact proof for the existence of the ground state
consisted of the rung-dimerized state and the ground state of the
decoupled chain. The phase diagrams of the spin-$1/2$ and spin-$1$
tetrahedron chain are given and the phase boundaries are precisely
determined for the small-size systems. For both the spin-$1/2$ and
spin-$1$ systems, there exist three phases, say, the fully
dimerized singlets on the rungs plus a decoupled chain, a mixed
phase with alternating spin singlet and state with total rung spin
$2S$ on the rungs and a ferrimagnetic phase with long-range order,
as the strength of vertical exchanges varies from infinity to
minus infinity. We also studied the $SU(3)$ spin-exchange model on
the 1D tetrahedron chain, for which the ground sate is a double
degenerate trimerized state and the elementary excitations are
fractionalized topological excitations. Our results indicate that
the properties of the ground state for the pyrochlore systems with
half-integer and integer
spins or the systems with the same spins but different internal symmetries ($%
SU(2)$ and $SU(3)$ for spin-1 systems) are quite different.

\begin{acknowledgments}
S. C. would like to acknowledge hospitality of the physical
department of the Chinese University of Hongkong during his
visiting period and he also thanks the Chinese Academy of Sciences
for financial support. This work is supported in part by NSF of
China under Grant No. 10574150 and Grant No. 10329403, and RGC
401504. C. W. is supported by the NSF under the Grant No.
Phy99-07949.
\end{acknowledgments}

\end{document}